\documentclass[superscriptaddress,showkeys,twocolumn,nofootinbib,longbibliography]{revtex4-2}

\usepackage{amsmath}
\usepackage{amssymb}
\usepackage{graphicx}
\usepackage{epsf}
\usepackage{slashed}
\usepackage{enumitem}
\usepackage[czech,english]{babel}
\usepackage[cp1250]{inputenc}
\usepackage{hyperref}
\usepackage{xcolor}

\usepackage[only,llbracket,rrbracket,llparenthesis,rrparenthesis]{stmaryrd} % for comparison with 4 similar parenthesis symbols
\usepackage{accsupp} % for ensuring the right Unicode codepoint upon pasting

\newcommand{\refer}[1]{(\ref{#1})}

\usepackage{bbm} %\mathbbm{1}, \mathbbmss{1}, \mathbbmtt{1}

\usepackage[nodayofweek]{datetime}
\newdateformat{mydate}{\twodigit{\THEDAY}{ }\shortmonthname[\THEMONTH], \THEYEAR}
\usepackage[normalem]{ulem}
\begin{document}

\title{Non-adiabatic theory of the hydrogen bond. Quantum computation?}

\author{I. Huba\v{c}}
\email{belaxx@gmail.com}
\affiliation{Institute of Physics and Research Centre of Theoretical Physics and Astrophysics, Silesian University in Opava, Bezru\v{c}ovo n\'am. 1150/13, 746 01
 Opava, Czech Republic}
 
\author{S. Wilson }
\email{'This paper is dedicated to the memory of  Dr. Stephen Wilson  DSC who passed away on 2nd September, 2020. He played an instrumental role in the initial stages of the study presented here.}
\affiliation{The Physical and Theoretical Chemistry Laboratory, University of Oxford}

\begin{abstract}
The hydrogen bond is usually described within the clamped nucleus approximation in which electronic and vibrational motions are considered separately. This approach leads to a double-well potential which facilitates proton tunnelling. In this work, the hydrogen bond is described by a formalism based on the non-adiabatic Hamiltonian in which electronic and vibrational motions are coupled. Quasi-degeneracy associated with the hydrogen bond supports this approach which is shown to afford an alternative picture of proton tunnelling.\\
\end{abstract}


\maketitle

%\tableofcontents

\section{Introduction}

The importance of the hydrogen bond in chemistry and molecular biology cannot be over-emphasized. For example, hydrogen bonding supports an efficient and versatile bonding mechanism for biological macromolecules since, although weak in comparison with covalent and other types of bonds, large numbers of hydrogen bonds can stabilize macromolecular systems without reducing flexibility. In 2011, a IUPAC Task Group (Physical and Biophysical Chemistry Division) \cite{arunan, arunan2} recommended a modern evidence-based definition of hydrogen bonding.

\begin{quotation}
\emph{`The hydrogen bond is an attractive interaction between a hydrogen atom from a molecule or a molecular fragment XH in which X is more electronegative than H, and an atom or a group of atoms in the same or a different molecule, in which there is evidence of bond formation.'}
\end{quotation}

The Task Group state a \emph{`typical hydrogen bond may be depicted as XH ... YZ, where the three dots denote the bond.'} Implicit in this picture of the hydrogen bond is the assumption of the separation of electronic and nuclear motion contained in the clamped-nucleus or Born-Oppenheimer approximation \cite{born, born2, born3, blinder, kolos, sutcliffe}; an approximation which underpins the concept of molecular structure \cite{sutcliffe2}. The iupac paper \cite{arunan2} recognizes that \emph{`One of the most powerful tools to study hydrogen-bonded systems, or intermolecular interactions in general, is computational chemistry.'} Typical examples can be found in an extensive literature, e.g. \cite{peterson, scheiner, frey, tauer}. Molecular electronic structure theory \cite{wilson} is almost invariably based on the clamped-nucleus approximation which in turn is dependent \emph{`on the large ratio between electronic and nuclear masses'} \cite{weeny}. The normal hydrogen atom nucleus, i.e. a proton, is the lightest of nuclei and therefore its motion is more susceptible to coupling with the motions of electrons than other, heavier atoms. 

The clamped-nucleus treatment of the hydrogen bond leads to a double-well potential generated by the electrons in which motion of the proton takes place. The double-well potential allows proton tunnelling. 

Electrons are fermions. They are indistinguishable. They can be described by quantum mechanical many-body theories. 

Nuclei can be treated through their vibrational motion. The quantum mechanical description of the nuclei in a molecule and their interactions with the electrons can be achieved by introducing a quasi-particle, the phonon. The phonon is widely used in the study of condensed matter systems. The phonon is associated with the vibrations of the nuclei about some assumed equilibrium positions. Our previous work described the general theory of electron-phonon interactions in molecules \cite{rag1}. Here we describe the application of this theoretical apparatus to hydrogen bonds.

In the present study, the \emph{non-adiabatic} theory of the hydrogen bond is considered. It involves the quantum mechanical treatment of coupled electronic and nuclear motion. The key idea is to use effective quasiparticles that consist of electrons dressed in \emph{phonons}. This construction lends itself naturally to an image of a Kitaev-like chain of quasiparticles that is a source of delocalized Majorana fermions.  In this way, a pair of hydrogen bonds (such as A-T base pair) may be endowed with a single quantum bit. Hence, we point out a possible role of hydrogen bonds as facilitators of quantum computation that may take place in DNA.

\section{The clamped-nucleus approximation for hydrogen bonds}

The separation of electronic and nuclear motion in molecular systems is ubiquitous in quantum chemistry. This clamped-nucleus or Born-Oppenheimer approximation \cite{born, born2, born3, blinder, sutcliffe} underpins the concept of molecular structure \cite{sutcliffe2}. Physically, because of the large disparity of the electron and nuclear masses, the electrons are said to adapt instantaneous to any change in the positions of the nuclei. The electrons generate an effective potential in which the nuclear motion takes place. 

The electronic structure problem \cite{wilson} can be attacked by exploiting many-body methods. The many-body problem \cite{wilson2} is attacked first by solving a mean-field model in which the averaged interactions of the electrons are described and then by developing a series of corrections, \emph{i.e.} electron correlation. The total molecular electronic Hamiltonian for a system of M nuclei and N electrons is written as
\begin{equation}
\mathcal{H}_{e} = \sum_{p=1}^{N} h_{p} + \sum_{p>q}^{N} g_{pq}\,,
\end{equation}
where the one-electron term is the sum of a term describing the kinetic energy of the
electrons and a term arising from the Coulomb attraction between the electrons and the nuclei:
\begin{equation}
h_{p} = - \frac{1}{2}\triangledown_{p}^{2} + \sum_{A=1}^{M} \frac{Z_{A}}{r_{pA}}\,.
\end{equation}
The two-electron term
\begin{equation}
g_{pq} = \frac{1}{r_{pq}}
\end{equation}
is associated with the Coulomb repulsions between the electrons. Independent electron models employ an N-electron Hamiltonian of the form:
\begin{equation}
\mathcal{H}_{\rm eff} = \sum_{p=1}^{N} (h_{p} + u_{p})\,,
\end{equation}
where $ u_{p} $ is a one-electron, mean-field potential. 

The $N$-electron Hamiltonian is written as a sum of $N$ one-electron Hamiltonians, while $ u_{p} $ provides a description of the averaged interactions between the electrons and also the screening of the nuclear charges. The total molecular electronic Hamiltonian can be written as
\begin{equation}
\mathcal{H}_{e} = \mathcal{H}_{\rm eff} + \left\{\sum_{p>q}^{N} g_{pq} - \sum_{p}^{N} u_{p}\right\}
\end{equation}
The terms in parenthesis, which are dubbed the `fluctuation potential', will be small if the independent electron model is well chosen. Much progress was achieved by employing second quantization formalism. Electronic Hamiltonian can be written in the so-called normal product (N product) form:
\begin{eqnarray}
\mathcal{H}_{e} = &\langle \phi_{0}|H|\phi_{0}\rangle + \sum_{i_{1},i_{2}} \langle i_{1}|f|i_{2}\rangle N[a_{i_{1}}^{+}a_{i_{2}}]\\
&+ \frac{1}{2} \sum_{i_{1}i_{2}i_{3}i_{4}}\langle i_{1}i_{2}|g|i_{3}i_{4}\rangle N[a_{i_{1}}^{+}a_{i_{2}}^{+}a_{i_{4}}a_{i_{3}}]\nonumber
\end{eqnarray}

This led to diagrammatic many-body Rayleigh-Schr$\ddot{o}$dinger perturbation theory (MBPT) and coupled cluster (CC) methods \cite{paldus}.

\section{Non-adiabatic Hamiltonian and quasiparticle description of hydrogen bonds}

A description of molecular systems which goes beyond the clamped-nucleus approximation involves the quantum mechanical treatment of both electronic and nuclear motion \cite{matyus}. Detail treatment of separation of electronic and nuclear motion can be found in the book by Born and Huang (see Appendix VIII) \cite{born3}.

According to Born and Huang \cite{born3} when spacing between two minima  $ |E_{i}-E_{j}| $ is comparable to vibrational quantum $\hbar \omega$, we cannot use Born-Oppenheimer (BO) approximation. Specifically, the condition  for using BO approximation is 
\begin{equation}\label{eq:condbo}
\frac{\hbar \omega}{|E_{i}-E_{j}|}\ll 1 \,. 
\end{equation}
In case when this condition is not valid we have to take into consideration also the coupling of electronic and vibrational motions. 

\begin{figure}
\includegraphics[width=\linewidth]{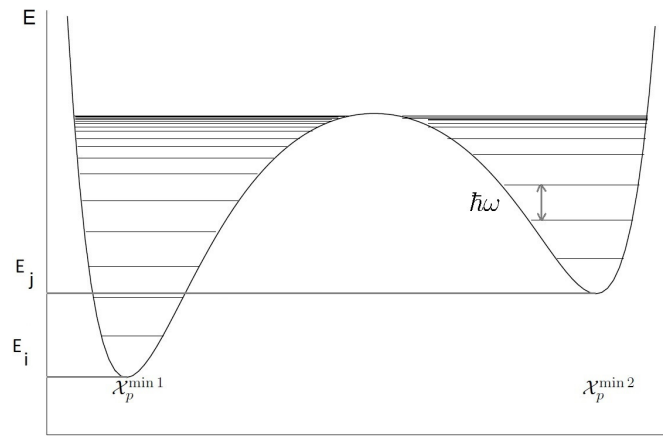}
\caption{Schematic representation of a hydrogen bond in the Born-Oppenheimer approximation}
\label{fig:01}
\end{figure}

Double-well H-bond potential in DNA is quasi-degenerate (see Fig.\ref{fig:01}). The difference between energy levels of both minima being comparable to $\hbar \omega$ puts us precisely into the situation where validity of BO approximation is questionable.

In our previous work \cite{svrcek1, svrcek2, svrcek3} we developed non-adiabatic molecular Hamiltonian. It is based on second quantization and mixing fermion and boson statistic. This is quasi-particle concept. We have also shown that our molecular non-adiabatic  Hamiltonian is fully equivalent with non-adiabatic Hamiltonian used in solid state physics.  The creation and annihilation operators from electronic Hamiltonian are combined with creation and annihilation operators of harmonic oscillator. The basic idea is the construction of fermion creation and annihilation operators as depending on normal coordinate and corresponding momentum of harmonic oscillator. 

The quasiparticle transformation relating the crude adiabatic representation with our new quasiparticle representation  may then be written as follows:
\begin{eqnarray}
\bar{a}_{P} &= a_{P} + \sum\limits_{Q}\sum\limits_{k=1}^{\infty} \frac{1}{k!} \sum\limits_{r_{1}\dots r_{k}} C_{PQ}^{r_{1}\dots r_{k}} B_{r_{1}} \dots B_{r_{k}} a_{Q} \,,   \\
\bar{a}_{P}^{\dagger} &= a_{P}^{\dagger} + \sum\limits_{Q}\sum\limits_{k=1}^{\infty} \frac{1}{k!} \sum\limits_{r_{1}\dots r_{k}} C_{PQ}^{r_{1}\dots r_{k}} B_{r_{1}} \dots B_{r_{k}} a_{Q}^{\dagger}  \,, 
\end{eqnarray}
where $B_{r_i}$ are normal coordinates and $C_{PQ}$ are expansion coefficients. In second quantization, normal coordinates $B$ and corresponding momentum $\widetilde{B}$ are defined as: 
\begin{eqnarray}
B&=b+b^{\dagger} \,,\\
\widetilde{B} &= b-b^{\dagger} \,.
\end{eqnarray}

Vibrational Hamiltonian $H_{\rm vibr}$ is given as:
\begin{eqnarray}
H_{\rm vibr}&=\sum\limits_{r} \hbar \omega_{r} \left ( b_{r}^{\dagger}b_{r} + \frac{1}{2} \right ) \,,
\end{eqnarray}
 where $b_{r}^{\dagger}$ and $b_{r}$ are bosonic creation and annihilation operators, respectively. 
 Schematically, quasiparticle transformation is facilitated by Bogoliubov-like transformation:
\begin{eqnarray}
\bar{a}_{P} &= \sum_{Q} C_{PQ}(B)a_{Q} \,,\\
\bar{a}_{P}^{\dagger} &= \sum_{Q} C_{PQ}(B)^{\dagger}a_{Q}^{\dagger} \,.
\end{eqnarray}

This can be generalized to also include momenta as \cite{svrcek4}
\begin{eqnarray}
\bar{a}_{P} &= \sum\limits_{Q} C_{PQ}(B, \tilde{B})a_{Q}\,,   \label{eqn:a1} \\
\bar{a}_{P}^{\dagger} &= \sum\limits_{Q} C_{PQ}(B, \tilde{B})^{\dagger}a_{Q}^{\dagger} \,.  \label{eqn:a2}
\end{eqnarray}

\begin{figure}
\includegraphics[width=\linewidth]{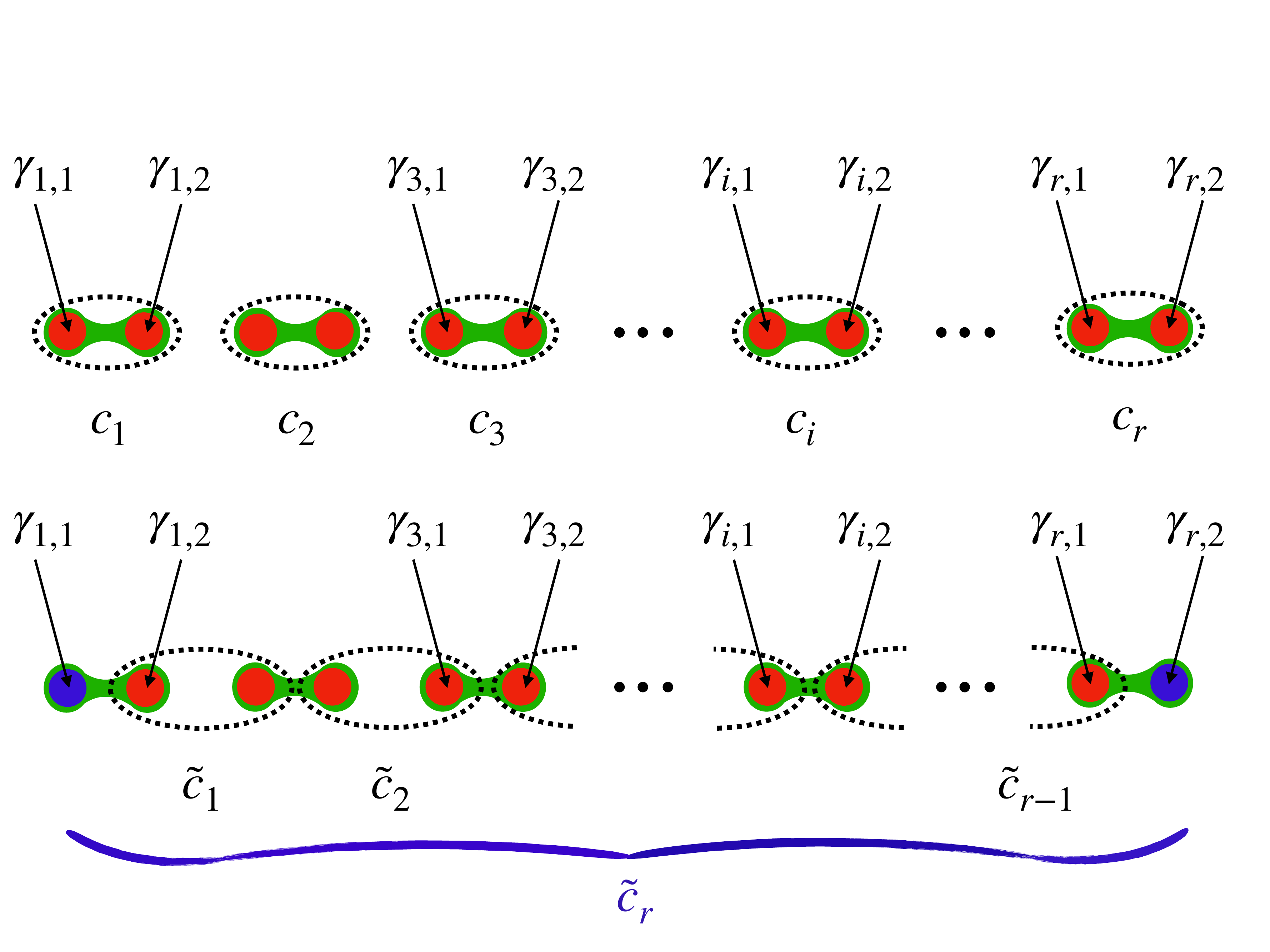}
\caption{A Kitaev chain. In product of fermion operators $ c_{1}, c_{2} ... c_{i} ... c_{r} $ each operator is written as a sum of two Majorana operators $ \gamma_{i,1} $ and $ \gamma_{i,2} $. The Majorana operators are then recoupled so as to leave the isolated $ \gamma $ operators  $ \gamma_{i,1} $ and  $ \gamma_{i,2} $ at either end of the chain. (The notation of Leijnse and Flensberg \cite{leijnse})  }
\label{fig_kitaev_chain}
\end{figure}

Note that this quasiparticle transformations are completely general without any departure from exactness, provided that we can specify coefficients $C_{PQ}$. Without going into details, however, let us now apply the quasiparticle description to H-bonds to develop the following intuition. Since the electrons directly couples to vibrations (position and momentum) of the proton that is delocalized along the hydrogen bond, \emph{the phonon-dressed fermions ($\bar{a}_P$,$\bar{a}_P^{\dagger}$) are thus linearly delocalized on the hydrogen bond as well}. We propose that due to this dependence the quasi-particles forms a one-dimensional chain along the bond that is very similar to a Kitaev chain \cite{leijnse}. The key ingredient is the effective one-dimensionality. It is known that Kitaev chain represent a stable coherent state. Formally, this is described using effective Majorana operators into which any fermionic operator can be decomposed. Schematically:
%Changing variables step by step in eq. (\ref{eqn:a1}).and (\ref{eqn:a2}) leads to chain of fermion quasiparticles. This chain fulfills the form of Kitaev chain \cite{leijnse} see Fig.\ref{fig_kitaev_chain}, since every fermion can be rewritten as combination of Majorana. We can put 
\begin{eqnarray}
\bar{a}_{P} &= \gamma_{P,1 } + i \gamma_{P, 2}\,, \\
\bar{a}_{P}^{\dagger} &= \gamma_{P, 1} - i \gamma_{P, 2}\,,
\end{eqnarray}
where $ \gamma_{P, 1} $ and $ \gamma_{P, 2} $ are Majorana operators. The edge state generated by a pair of Majorana operators is delocalized and thus stable agains local decoherence (see Fig.~\ref{fig_kitaev_chain}).

%This concept we apply on double well H-bond potential in DNA. 
%In this way we can look on hydrogen bonds as constructed by quasiparticles forming Kitaev-like chains \cite{leijnse, rag2}.% Since these H-bonds are one-dimensional, they resembles and fulfill the form of Kitaev chain of Majorana quasiparticles \cite{leijnse, rag2}. 

\section{Quantum computation on H-bond of DNA}

Let us now assume that our intuitive picture of H-bonds is correct and let us see what that could imply for the nature of quantum computation that may be taking place in DNA.

Since A-T  base pairs involve two H-bonds, we have four Majorana states on their edges. 
 %Thus A-T ( T-A) correspond to four Majorana. 
 Four Majorana form one q-bit.
On the other hand, one C-G base pair has three H-bonds and yields three q-bits that form an entangled state with quantum entropy (see \cite{rag1}). 
 %Kitaev chain represent a stable coherent state.  Decoherence  will destabilize this state. 
 In our RAGtime \cite{rag2} paper, we studied how effective dynamics of DNA strand may lead to decoherence\footnote{This is a \emph{configurational decoherence}, not a decoherence in a traditional sense that concerns a stability of a quantum information due to environment.} of H-bonds, resulting in braiding \cite{leijnse}. Specifically, 
in \cite{rag2} paper, we studied the effect of collision of 2 solitons on double-well H-bonds in DNA. In this way, we simulated de-coherence,
however, the particular mechanism of the decoherence is not important here. In line with the arguments of L\"owdin \cite{lowdin1, lowdin2}, we can envisage the braiding on q-bits A-T (T-A) as we set out below. 
%taking into account that 4 Majorana form one q-bit and we shall follow Lowdin`s way \cite{lowdin1, lowdin2} of  proton tunnelling on double well H-bond.

As we have discussed \cite{rag1}, we treat double-well H-bond potential in  DNA as a non-adiabatic system. Within this, we apply quasiparticle transformation of non-adiabatic Hamiltonians \cite{rag1}. This concept leads us to Majorana fermions and Kitaev chains.

Owing to some dynamics, let us assume that the bond distance between A and T molecules becomes temporarily shorter. This changes the coupling between the electronic and the vibrational motions, which is the form of  decoherence. As a result, the double-well potential becomes closer to a single-well one or the difference between two minima change. Either way, the condition for the applicability of the BO approximation \refer{eq:condbo} will be again fulfilled. The protom may change its position along the potential energy curve. After that, as the source of de-coherence ceases, the phonon-electron coupling becomes again dominant and we return to non-adiabatic quasiparticle description. However, the value of the q-bit on the hydrogen bond has been flipped. This is the braiding of A - T into ${\rm A}^{*}$ - ${\rm T}^{*}$  systems. Schematically, initial states of the two H-bonds
% The braiding for A-T (and T-A) systems is as follows. We use the same argument for proton shift as was used by Lowdin \cite{lowdin1,lowdin2}.

\begin{figure}[h]
\includegraphics[width=\linewidth]{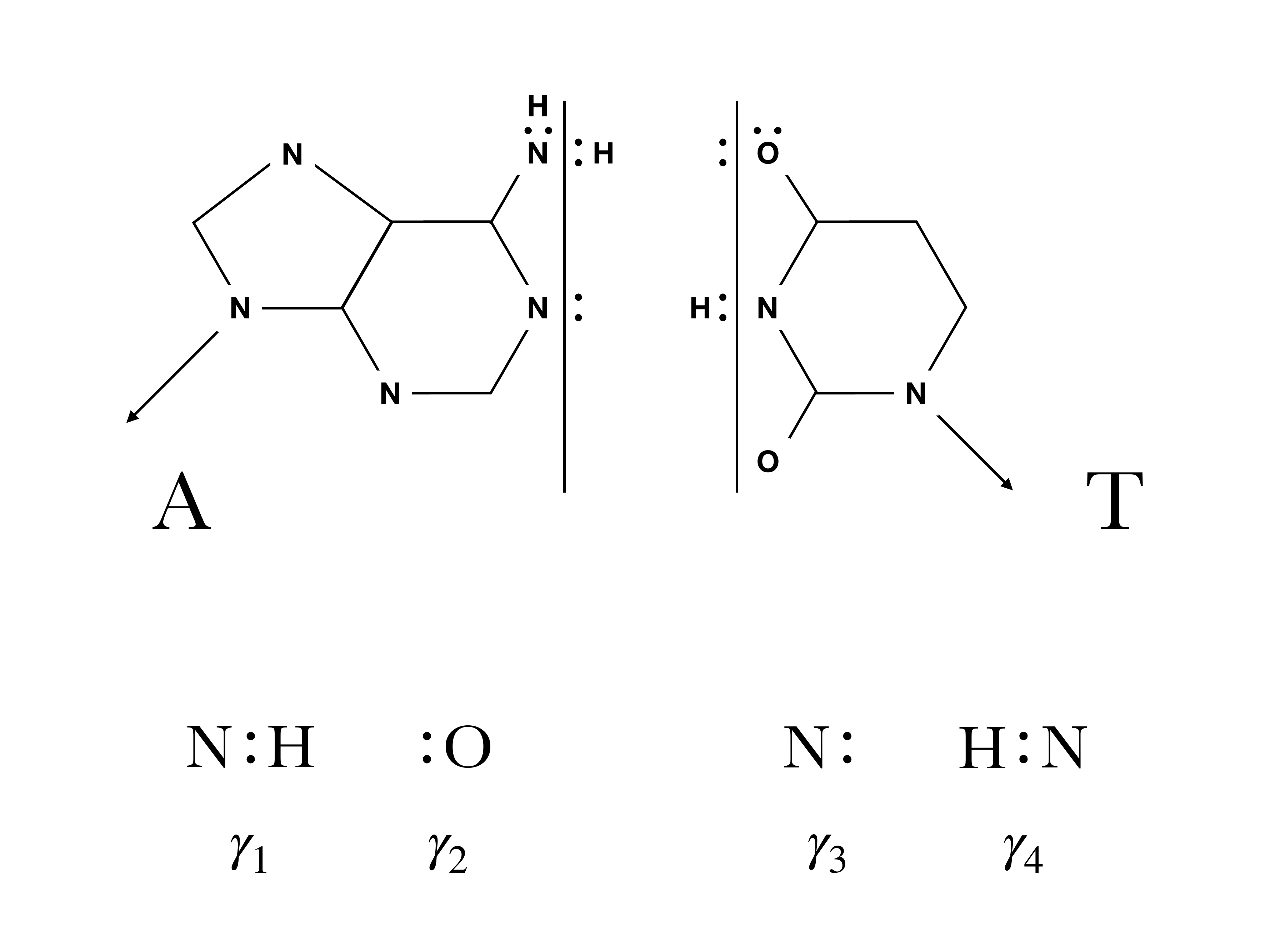}
\end{figure}

%\begin{equation*}
%\begin{array}{cccccccc}
%\mathrm{N}:  \mathrm{H} &&: \mathrm{O}  &  & &\mathrm{N}: & & \mathrm{H}:\mathrm{N} \\

%\mathrm{\gamma_{1}}&  & \mathrm{\gamma_{2}} & & & \mathrm{\gamma_{3}} && \mathrm{\gamma_{4}}
%\end{array}
%\end{equation*}
are in this way transformed into:

\begin{figure}[h]
\includegraphics[width=\linewidth]{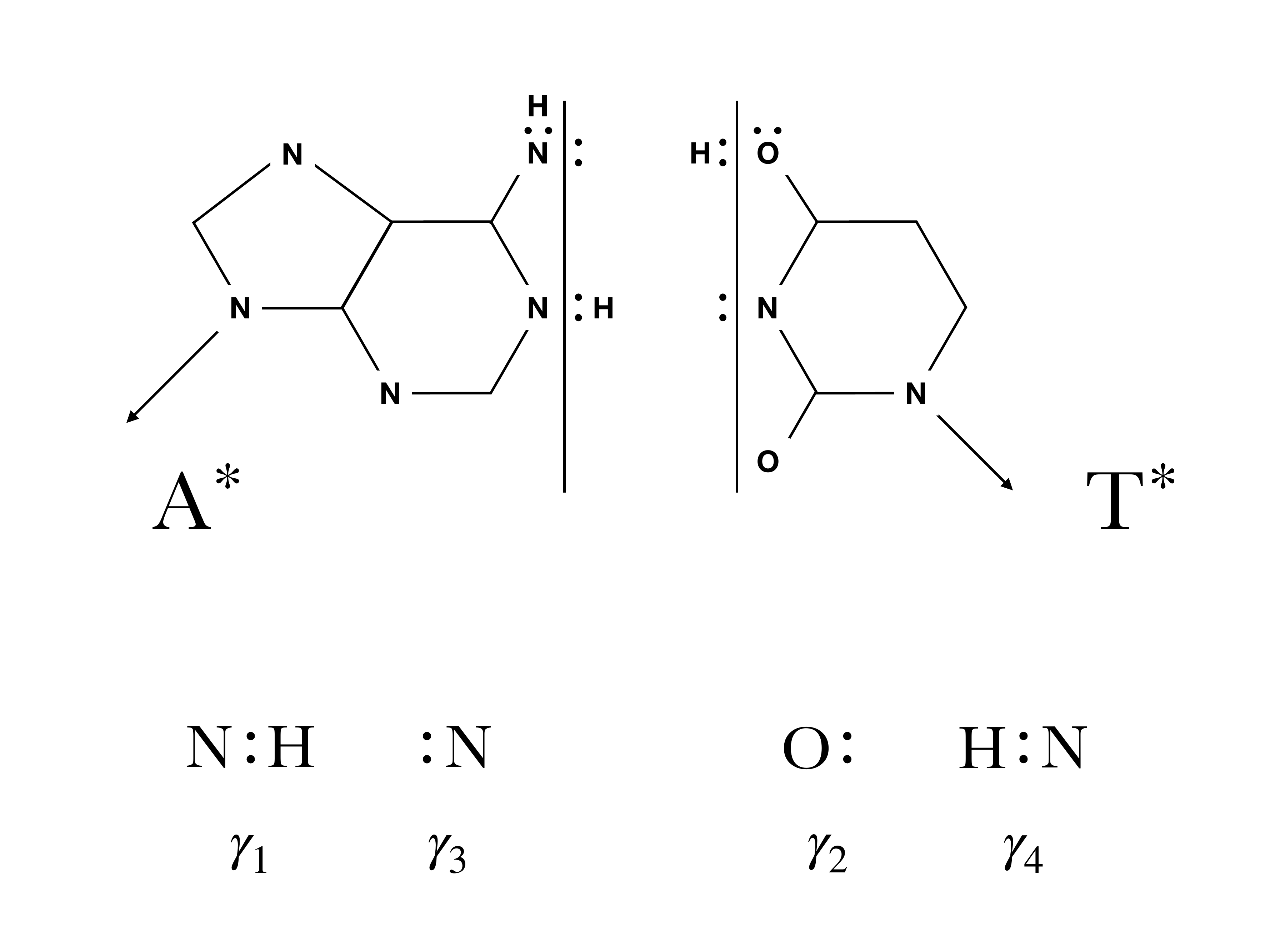}
\end{figure}

%\begin{equation*}
%\begin{array}{c c c c c}
%\mathrm{N}:\mathrm{H}\;\;\;\; \mathrm{N:}& &:\mathrm{O}  \;\;\;\;\mathrm{H}:\mathrm{N}\\
%\mathrm{\gamma_{1}}\;\;\;\;\;\;\;\;\mathrm{\gamma_{3}} &&\mathrm{\gamma_{2}}\;\;\;\;\;\;\mathrm{\gamma_{4}}
%\end{array}
%\end{equation*}
This is nothing but the braiding as an exchange of ($ \gamma_{1} \gamma_{2} \gamma_{3} \gamma_{4} $) into ($ \gamma_{1} \gamma_{3} \gamma_{2} \gamma_{4} $) (we follow the notation for A, T, ${\rm A}^{*}$ and ${\rm T}^{*}$  the notation of L\"owdin \cite{lowdin1,lowdin2}).

%\begin{eqnarray}
%N &: H   N :  : O  H : N\\
%&\gamma_{1}   \gamma_{3}   \gamma_{2} \gamma_{4}
%\end{eqnarray}

%\colvec{1}{0}

Therefore, due to braiding, we have (A - T) $ \rightarrow $ (${\rm A}^{*}$ - ${\rm T}^{*}$). If we assign (A - T) a q-bit $ \vert 0 \rangle = \left({1 \atop 0} \right)$, (${\rm A}^{*}$ - ${\rm T}^{*}$) q-bit is $ \vert 1 \rangle = \left({0 \atop 1} \right)$. This can lead to computation with one q-bit. 

Since one q-bit changes can be extended throughout the DNA we believe that it can play some role in correction of mutation in DNA or other mechanisms.

Of course, at this point it is impossible to say, whether our speculation is realized in Nature. A particularly unclear point is the stability of quantum information. If quantum computing is somehow undergoing in DNA, there must be some quantum error-correcting mechanism at play, especially in view of (or because?) mRNA transcription and DNA replication. However, as classic would say: if quantum computing is possible, it would be strange if Nature did not find a use for it.

\section{Conclusion}

The Born-Oppenheimer (BO) approximation leading to the separation of vibrational and electronic motion is the main tool used in quantum chemistry computational programs. The use of this approximation is based on the disparity between the mass of electron with respect to the nuclei.

As discussed by Born and Huang (Appendix VIII), there are situations when the use of this this is questionable. Particulary in the case of quasi-degeneracy in electronic states. Such a situation appears in double-well H-bond potential in DNA. According to Born and Huang, in such a case we have to take into account coupling between electronic and vibrational motion. 

BO approximation was applied on double-well H-bond potential by L\"owdin, who studied proton tunnelling on double-well potential. However, proton-tunnelling was not found even using most sophisticated quantum chemistry methods. 
This support our idea of quasiparticles based on coupling vibrational and electronic motion. Hence, it seems that L\"owdin's idea of tunnelling can be reconsidered as braiding quantum information on hydrogen bonds. 
Such an approach lead us to different interpretation of double-well H-bond potential. 
As discussed by L\"ovdin, quoting Delbr\"uck and Schr\"odinger \cite{schrodinger}:
\begin{quote}
`... These authors emphasized that there ought to be a close parallelism between the immense stability of the hereditary substance over thousands of years and the stationary state of giant molecule of `aperiodic solid,' and that further the discontinuous changes of the genetic code leading to mutations should correspond to `quantum jumps' between various stationary states. It seems now as if these quantum jumps would be associated with proton transfer within the nucleotide bases in the DNA molecule.' \cite{lowdin1}
\end{quote}

According to our idea, the stability of DNA is due to coherent state connected with Majorana fermions (Kitaev chain) on double-well H-bond potential. Discontinuous changes and quantum jumps are due to decoherence, braiding and change of q-bits (quantum information).
%Quasiparticle transformation permits us to introduce q-bits and to discuss one q-bit computation. This may play the role in correcting mutation on DNA.

\section*{Acknowledgement}
I.H.wants to thank Mrs. Kathryn Wilson, J. Moty\'{c}ka MSc, Dr. F. Blaschke  and O. N. Karp\'{i}\v{s}ek MSc  for help with preparing the manuscript.

%\section*{References}
\bibliographystyle{unsrt}
\bibliography{references}

\begin{thebibliography}{10}

\bibitem{arunan}
R.A. Klein J. Sadlej S. Scheiner I. Alkorta D.C. Clary R.H. Crabtree J.J.
  Dannenberg P. Hobza H.G. Kjaergaard A.C. Legon B. Mennucci D.J.~Nesbitt
  E.~Arunan, G.R.~Desiraju.
\newblock Definition of the hydrogen bond (iupac recommendations 2011).
\newblock {\em Pure Appl. Chem.}, 83, 1637, (2011).

\bibitem{arunan2}
R.A. Klein J. Sadlej S. Scheiner I. Alkorta D.C. Clary R.H. Crabtree J.J.
  Dannenberg P. Hobza H.G. Kjaergaard A.C. Legon B. Mennucci D.J.~Nesbitt
  E.~Arunan, G.R.~Desiraju.
\newblock Defining thr hydrogen bond: An account ( iupac technical report ).
\newblock {\em Pure Appl. Chem.}, 83, 1637, (2011).

\bibitem{born}
M.~Born and J.R. Oppenheimer.
\newblock Zur quantentheorie der molekuelen.
\newblock {\em Ann. Phys., Leipzig}, 84, 457, (1927).

\bibitem{born2}
Max Born.
\newblock {\em Die G{\"u}ltigkeitsgrenze der Theorie der idealen Kristalle und
  ihre {\"U}berwindung}, pages 1--16.
\newblock Springer Berlin Heidelberg, Berlin, Heidelberg, 1951.

\bibitem{born3}
M.~Born and K.~Huang.
\newblock {\em Dynamical theory of crystal lattices}.
\newblock Oxford University Press, (1954).

\bibitem{blinder}
S.M. Blinder.
\newblock {\em On the quantum theory of molecules}.
\newblock with emendations by Brian Sutcliffe and Wolf Geppert, published as an
  appendix to, (September 2001).

\bibitem{kolos}
W.~Kolos and L.~Wolniewicz.
\newblock Nonadiabatic theory for diatomic molecules and its application to the
  hydrogen molecule.
\newblock {\em Rev. Mod. Phys.}, 35, 473, (1963).

\bibitem{sutcliffe}
B.T. Sutcliffe.
\newblock {\em Handbook of Molecular Physics and Quantum Chemistry Fundamentals
  599}.
\newblock John Wiley, Chichester, (2003).

\bibitem{sutcliffe2}
B.T. Sutcliffe.
\newblock {\em Handbook of Molecular Physics and Quantum Chemistry Fundamentals
  588}.
\newblock John Wiley, Chichester, (2003).

\bibitem{peterson}
T.H.~Dunning K.A.~Peterson.
\newblock Benchmark calculations with correlated molecular wave functions. vii.
  the structure and binding energy of the hf dimer.
\newblock {\em J. Chem. Phys.}, 102, 2032, (1995).

\bibitem{scheiner}
S.~Scheiner.
\newblock Ab initio studies of hydrogen bonds: The water dimer paradigm.
\newblock {\em Annu. Rev. Phys. Chem.}, 45, 23, (1994).

\bibitem{frey}
S.~Leutwyler J.A.~Frey.
\newblock An ab initio benchmark study of hydrogen bonded formamide dimers.
\newblock {\em J. Phys. Chem. A}, 110, 12512, (2006).

\bibitem{tauer}
C.D.~Sherrill T.P.~Tauer, M.E.~Derrick.
\newblock Estimates of the ab initio limit for sulfur-$\pi$ interactions: The
  h$_{2}$s-benzene dimer.
\newblock {\em J. Phys. Chem. A}, 109, 191, (2005).

\bibitem{wilson}
P.F.~Bernath S.~Wilson and R.~McWeeny (eds.).
\newblock {\em Handbook of Molecular Physics and Quantum Chemistry 2 Molecular
  Electronic Structure}.
\newblock John Wiley, Chichester, (2003).

\bibitem{weeny}
R.~McWeeny.
\newblock {\em Methods of Molecular Quantum Mechanics}.
\newblock Academic Press, London, (1989).

\bibitem{rag1}
Miloslav~\v{S}vec Ivan~Huba\v{c} and Stephen Wilson.
\newblock Quantum entanglement and quantum information in biological systems
  (dna).
\newblock {\em Proceedings of RAGtime}, 19, 1 - 5 Nov., (2017).

\bibitem{wilson2}
S.~Wilson.
\newblock {\em Electron correlation in molecules}.
\newblock Clarendon Press, Oxford, (1984).

\bibitem{paldus}
J.~Paldus and \v{C}\'{i}\v{z}ek.
\newblock Time-independent diagrammatic approach to perturbation theory of
  fermion systems.
\newblock {\em Adv. Quantum Chem.}, 9, 105-97, (1975)).

\bibitem{matyus}
E.~M\'{a}tyus.
\newblock Pre-born-oppenheimer molecular structure theory.
\newblock {\em Molec. Phys.}, 117, 590609, (2019).

\bibitem{svrcek1}
I.~Huba\v{c} and M.~Svr\v{c}ek.
\newblock The quasiparticle concept in vibrational-electronic problems in
  molecules.
\newblock {\em Rev. Modern Phys}, 33, pp. 403-443, ISSN 1097-461X, (1988)).

\bibitem{svrcek2}
I.~Huba\v{c} and M.~Svr\v{c}ek.
\newblock {\em Many-Body Perturbation Theory for Vibrational Electronic
  Molecular Hamiltonian, in S.Wilson and G. H. F. Diercksen, editors,}, volume
  293, pp. 471-512, ISBN 978-1-4615-7419-4.
\newblock NATO ASI Series (Series B: Physics), Springer, Boston, (1992)).

\bibitem{svrcek3}
I.~Huba\v{c} and M.~Svr\v{c}ek.
\newblock {\em Many-Body Perturbation Theory for Vibrational Electronic
  Molecular Hamiltonian, in S.Wilson and G. H. F. Diercksen, editors,}, volume
  4, Molecular vibrations.
\newblock Springer US, ISBN 9780306441684, (1992)).

\bibitem{svrcek4}
M.~Svr\v{c}ek.
\newblock Ph.d. theis.
\newblock {\em Faculty of Mathematics and Physics, Comenius University,
  Bratislava}, (1986).

\bibitem{leijnse}
M.~Leijnse and K.~Flensberg.
\newblock {\em Semiconductor Science and Technology}, volume~27.
\newblock 2012.

\bibitem{rag2}
O.~Nicolas Karp\'{i}\v{s}ek M.~\v{S}vec I.~Huba\v{c}, F.~Blaschke and
  S.~Wilson.
\newblock Quantum information in biomolecules: transcription and replication of
  dna using a soliton model.
\newblock {\em Proceedings of RAGtime}, 22, 19-23 Oct, (2020).

\bibitem{lowdin1}
Per-Olov L\"owdin.
\newblock The normal constants of motion in quantum mechanics treated by
  projection technique.
\newblock {\em Rev. Mod. Phys.}, 34:520--530, Jul 1962.

\bibitem{lowdin2}
Per-Olov L\"owdin.
\newblock Advances in quantum chemistry, volume 2, academic press, 1st edition.
\newblock 2, (1966).

\bibitem{schrodinger}
Erwin Schr\"odinger.
\newblock {\em What is Life? The Physical Aspect of the Living Cell}.
\newblock Cambridge University Press, 1944.

\end{thebibliography}

\end{document}